\documentclass[]{iac}
\usepackage[acronym]{glossaries}
\usepackage{hyperref}
\usepackage[citestyle=numeric,bibstyle=numeric,maxbibnames=1,maxcitenames=1,sorting=none,sortcites]{biblatex}
\usepackage[dvipsnames,table,xcdraw]{xcolor}
\usepackage[inline]{enumitem}
\usepackage{siunitx}
\usepackage[nameinlink]{cleveref}
\usepackage{tabularx}
\usepackage{acro}
\usepackage[british]{babel}
\usepackage{xcolor-material}
\usepackage{soul}
\usepackage{datetime}
\usepackage{float}
\usepackage{caption}
\usepackage{subcaption}
\usepackage{cleveref}
\usepackage{multicol}
\usepackage{tabularx}
\usepackage{gensymb}

\definecolor{draft_color}{HTML}{FF0000}

\newdate{campaign_start_1}{05}{03}{2023}
\newdate{campaign_finish_1}{09}{03}{2023}
\newdate{campaign_start_2}{15}{04}{2024}
\newdate{campaign_finish_2}{19}{04}{2024}
\acsetup{use-id-as-short}

\DeclareAcronym{CDS}{long = CubeSat Design Specification}
\DeclareAcronym{TC}{long = Telecommand}
\DeclareAcronym{TM}{long = Telemetry}
\DeclareAcronym{DUT}{long = Device Under Test}
\DeclareAcronym{EPS}{long = Electrical Power System}
\DeclareAcronym{EQM}{long = Engineering Qualification Model}
\DeclareAcronym{TVAC}{long = Thermal Vacuum Chamber}
\DeclareAcronym{CDR}{long = Critical Design Review}
\DeclareAcronym{IC}{long = Integrated Circuit}
\DeclareAcronym{OBC}{long = On Board Computer}
\DeclareAcronym{AIV}{long = {Assembly, Integration and Verification}}
\DeclareAcronym{TRP}{long = Temperature Reference Point}
\DeclareAcronym{PCB}{long = Printed Circuit Board}
\DeclareAcronym{ADCS}{long = {Attitude, Determination and Control Subsystem}}
\DeclareAcronym{FM}{long = Flight Model}
\DeclareAcronym{AUTh}{long = Aristotle University of Thessaloniki}
\DeclareAcronym{ESA}{long = European Space Agency}
\DeclareAcronym{NCR}{long = Non - Conformance Report}
\DeclareAcronym{RC}{long = Resistor Capacitor}
\DeclareAcronym{CAN}{long = Control Area Network}
\DeclareAcronym{MCU}{long = Microcontroller Unit}
\DeclareAcronym{VIBE}{long = Vibration}
\DeclareAcronym{TCXO}{long = Temperature - Compensated Crystal Oscillator}
\DeclareAcronym{DRD}{long = Document Requirements Definition}
\DeclareAcronym{FT}{long = Functional Test}
\DeclareAcronym{VCD}{long = Verification Control Document}
\DeclareAcronym{ECSS}{long = European Cooperation for Space Standardization}
\DeclareAcronym{MRAM}{long = Magnetoresistive Random Access Memory}
\DeclareAcronym{LED}{long = Light-Emitting Diode}
\DeclareAcronym{STL}{long = C++ Standard Template Library}
\DeclareAcronym{ONFI}{long = Open NAND Flash Interface}
\DeclareAcronym{I2C}{long = Inter-integrated Circuit}
\DeclareAcronym{FDIR}{long = Fault Detection Isolation and Recovery}
\DeclareAcronym{SEE}{long = Single-Event Effect}
\DeclareAcronym{SEU}{long = Single-Event Upset}
\DeclareAcronym{SET}{long = Single-Event Transient}
\DeclareAcronym{SEL}{long = Single-Event Latch-up}
\DeclareAcronym{ESD}{long = Electrostatic Discharge}
\DeclareAcronym{RT}{long = Radiation Tolerant}
\DeclareAcronym{CAN-FD}{short = CAN-FD, long = CAN Flexible Data Rate}
\DeclareAcronym{PUS}{long = Packet Utilization Standard}
\DeclareAcronym{LCL}{long = Latch-up Current Limiter}
\DeclareAcronym{SATNOGS}{short = SatNOGS, long = Satellite Networked Open Ground Station}
\DeclareAcronym{ETL}{long = Embedded Template Library}
\DeclareAcronym{GSE}{long = Ground Support Equipment}
\DeclareAcronym{SMC}{long = Static Memory Controller}
\DeclareAcronym{NAND}{long = NOT-AND}
\DeclareAcronym{RTOS}{long = Real Time Operating System}
\DeclareAcronym{ARM}{long = Advanced RISC Machines}
\DeclareAcronym{TID}{long = Total Ionizing Dose}
\DeclareAcronym{CAD}{long = Computer-aided design}
\DeclareAcronym{HI}{long = Heavy-Ion}
\DeclareAcronym{EM}{long = Environmental Model}
\DeclareAcronym{MTQs}{long = Magnetorquers}
\DeclareAcronym{COMMS}{long = Communications}
\DeclareAcronym{EGSE}{long = Electrical Ground Support Equipment}
\DeclareAcronym{LoC}{long = Lab-on-a-Chip}
\DeclareAcronym{COTS}{long = Commercial of the Shell}
\DeclareAcronym{RDM}{long = Radiation Design Margin}
\DeclareAcronym{MAIVP}{long = Manufacturing Assembly Integration and Verification Plan}
\DeclareAcronym{LEO}{long = Low Earth orbit}
\DeclareAcronym{ESEC}{long = European Space Security and Education Centre}
\DeclareAcronym{CSF}{long = CubeSat Support Facility}
\DeclareAcronym{EMC}{long = Electromagnetic Compatibility}
\DeclareAcronym{AOCS}{long = Attitude and Orbit Control System}
\DeclareAcronym{TSTP}{long = Test Specification \& Test Procedures}
\DeclareAcronym{IM}{long = Integration Model}
\DeclareAcronym{PFM}{long = Proto-flight Model}
\DeclareAcronym{DM}{long = Development Model}

\bibliography{references}

\usepackage[many]{tcolorbox}
\newtcbox{\todo}{
    on line,
    colback=red!5!white,
    colframe=red!75!black,
    coltitle=red!75!black,
    fonttitle=\bfseries,
    fontupper=\footnotesize,
    title=TODO,
    detach title,
    before upper={\tcbtitle\ },
    nobeforeafter,
    tcbox raise base,
    top=0pt,bottom=0pt,left=0mm,right=0mm,
    toprule=0mm,
    bottomrule=0mm,boxsep=0.7mm,
}

\def\todo#1{}

\DeclareSourcemap{
  \maps[datatype=bibtex]{
    \map[overwrite]{
      \step[fieldsource=doi, final]
      \step[fieldset=url, null]
      \step[fieldset=eprint, null]
    }  
  }
}

\newbibmacro{string+url}[1]{%
    \iffieldundef{url}{#1}{\href{\thefield{url}}{#1}}
}
\DeclareFieldFormat{title}{\usebibmacro{string+url}{\mkbibemph{#1}}}
\DeclareFieldFormat[article]{title}{\usebibmacro{string+url}{\mkbibquote{#1}}}
\DeclareFieldFormat[inproceedings]{title}{\usebibmacro{string+url}{\mkbibquote{#1}}}
\DeclareFieldFormat[thesis]{title}{\usebibmacro{string+url}{\mkbibquote{#1}}}

\newcommand\myshade{85}
\colorlet{mylinkcolor}{violet}
\colorlet{mycitecolor}{Turquoise}
\colorlet{myurlcolor}{Blue}

\hypersetup{
  linkcolor  = mylinkcolor!\myshade!black,
  citecolor  = mycitecolor!\myshade!black,
  urlcolor   = myurlcolor!\myshade!black,
  colorlinks = true,
}

\usepackage{array}
\newcolumntype{L}[1]{>{\raggedright\let\newline\\\arraybackslash\hspace{0pt}}m{#1}}
\newcolumntype{C}[1]{>{\centering\let\newline\\\arraybackslash\hspace{0pt}}m{#1}}
\newcolumntype{R}[1]{>{\raggedleft\let\newline\\\arraybackslash\hspace{0pt}}m{#1}}

\setlist{leftmargin=0.8cm}

\begin{document}

\lhead{}
\chead{\footnotesize 75\textsuperscript{th} International Astronautical Congress (IAC), Milan, Italy, 14-18 October 2024.\\ Copyright \copyright 2024 by the Aristotle University of Thessaloniki. Published by the IAF, with permission and released to the IAF to publish in all forms.}
\rhead{}

\makeatletter
\lfoot{IAC--\iac@paperyear--\iac@papernumber}\cfoot{}\rfoot{Page \thepage\ of \pageref{LastPage}}%
\makeatother

\IACpaperyear{24}
\IACpapernumber{E2,4,1,x89777}
\IACconference{75}
\IAClocation{Milan, Italy, 14-18 October 2023}
\IACcopyrightB{2024}{Aristotle University of Thessaloniki}

\title{The On-board Computer of the AcubeSAT mission}


\IACauthor{Konstantinos Tsoupos}{\url{ktsoupos@ece.auth.gr}}{1}
\IACauthor{Stylianos Tzelepis}{\url{tzelepisstelios@gmail.com}}{5}
\IACauthor{Georgios Sklavenitis}{\url{gsklaven@ece.auth.gr}}{1}
\IACauthor{Dimitrios~Stoupis}{\url{dstoupis@auth.gr}}{1, 2}
\IACauthor{Grigorios Pavlakis}{\url{grigpavl@ece.auth.gr}}{1}
\IACauthor{Panagiotis~Bountzioukas}{\url{panagiotis.bountzioukas@ext.esa.int}}{6, 1}
\IACauthor{Christina Athanasiadou}{\url{athanchris@ece.auth.gr}}{1}
\IACauthor{Lily Ha}{\url{lily.ha@ext.esa.int}}{4}
\IACauthor{David Palma}{\url{David.Palma@esa.int}}{3}
\IACauthor{Loris Franchi}{\url{Loris.Franchi@ext.esa.int}}{6}
\IACauthor{Prof. Alkis Hatzopoulos}{\url{alkis@ece.auth.gr}}{1}


\abstract{
\\
The architecture of the \acf{OBC} subsystem for CubeSats is naturally evolving,
characterized by the integration of low-power, lightweight components, cutting-edge microprocessors harnessing multiprocessor or heterogeneous computing, and sophisticated algorithms enabling autonomous operation in increasingly compact platforms.

\par Despite the ongoing standardization efforts within the space community, the unique payload requirements of the AcubeSAT mission, an open-source nanosatellite that aims to probe the effects of radiation and microgravity on eukaryotic cells in \acf{LEO} within the FYS3 program of the \acf{ESA} Education Office, necessitated the development of a custom, cost-effective module. This module shall not only meet the spatial and functional demands of the mission but also serve as an educational tool, providing valuable hands-on experience to project members involved in designing space-grade equipment and developing critical system software that follow the \acf{ECSS} Standard.

\par The single \acf{PCB} hosts two distinct subsystems: one resembling a conventional \acs{OBC} responsible for telecommand execution, telemetry fetching, onboard time synchronization, in-orbit patching, and \acf{FDIR}. The second entails the implementation of an \acf{AOCS} for both nominal and downlinking scenarios, facilitating a directional patch antenna for payload data transmission. The hardware for each subsystem resides on separate sides of the PCB and includes non-volatile memories for critical data and telemetry storage, custom radiation-tested \acf{LCL} protection circuits, sensors, interfaces with the Payload Subsystem, the actuators and the rest of the in-house and \acf{COTS} components, all while maintaining compatibility with the LibreCube standard. At the core of each subsystem lies a Radiation Tolerant \acf{ARM} Cortex-M7 MCU. This architecture not only decentralizes processing power, mitigating single points of failure but also leverages redundancy capabilities.

\par This paper aims to elucidate the decision-making process, design iterations, and development stages of the custom board and accompanying in-house software. Insights garnered from the initial partially successful environmental test campaign at the \acs{ESA} \acf{CSF} will be shared, along with the ensuing preparations, results, and lessons learned from subsequent testing endeavors in April 2024. Furthermore, the current developmental status will be discussed alongside future \acf{EMC} testing, integration plan on a FlatSat, and prospects for the open-source design as a cost-effective, and modular solution that can be tailored with little effort for upcoming missions.}

\maketitle

\acuseall

\section{Introduction}

\label{sec:mission description}
\begin{figure*}[!ht]
    \centering
    \includegraphics[width=\linewidth]{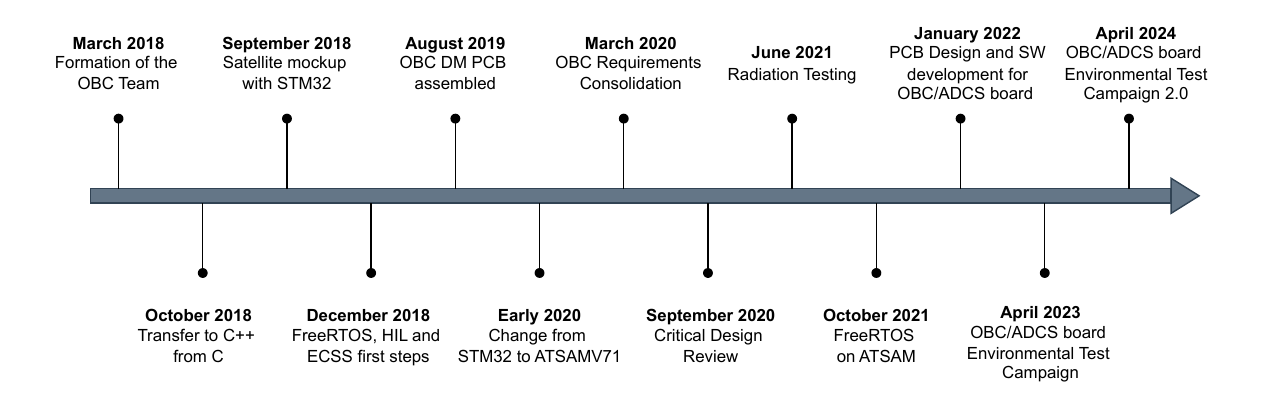}
    \caption{High-level timeline with the key milestones of the \acs{OBC} subsystem.}
    \label{fig:timeline}
\end{figure*}

AcubeSAT is an open-source Space biology project supported by the 3rd cycle of the Fly Your Satellite! programme of the \acs{ESA} Education Office, with a dual objective; first, it aims to probe changes in gene expression in eukaryotic cells when exposed to the effects of microgravity and radiation in Low-Earth Orbit. Second, it seeks to introduce a \acf{LoC} platform as a modular and cost-effective method for conducting high-throughput Space biology research. Beyond the scientific goals, SpaceDot, the team that is developing AcubeSAT, as an interdisciplinary Space research team of the Aristotle University of Thessaloniki, aspires to promote aerospace engineering as an accessible field of study within the academic community in Greece \cite{MDO}.

Throughout the development, the team focused on creating a custom On-Board Computer subsystem that would meet the unique needs of the mission, while ensuring reliability in the harsh Space environment \cite{langerReliabilityCubeSatsStatistical2016}. This paper outlines the journey, from the initial design stages to the rigorous testing of the PCB, highlighting key milestones, performed tests, and lessons learned along the way. A high-level timeline is illustrated in \Cref{fig:timeline}.

\section{Design Phase}

\subsection{History \& Background}
\label{sec:history}

Designing AcubeSAT’s custom \acs{OBC} marked one of the team’s first ventures into embedded systems development. The STM32L4S9ZIT6 \acf{MCU} from the STM32 series was initially chosen as the core of the subsystem, primarily due to the team’s familiarity with the platform and its suitable peripheral set that satisfied the mission requirements \cite{DDJF_SYS}. As reliability is paramount for a critical subsystem such as \acs{OBC}, a dual, cold-redundant \acs{MCU} setup was selected. 

However, due to space constraints and complexity the cold-redundant architecture was abandoned for a simpler design with a single \acs{MCU}. \emph{Microchip's (Atmel) SAMV71Q21RT} \cite{SAMV71Q21RT} was the new chip replacing the old STM32-based design, because of its low cost, resistance to radiation effects (as it has a drop-in pin-compatible radiation-tolerant version available), and similarity in the feature set with the previous \acs{MCU}. Since this design change, the SAMV71Q21RT has continued to serve as the primary processing unit for the subsystem up to the present day. This processor is also employed across some other in-house built subsystems, including the Payload \acs{PCB} and the \acf{ADCS}. By standardizing on the SAMV71Q21RT, decentralized processing power is achieved to meet the computational requirements of each subsystem individually, while also reducing the software development overhead and preserving robustness due to the radiation hardness of the \acs{MCU}.

\begin{figure*}[!hp]
    \centering
    \begin{subfigure}{0.495\linewidth}
        \centering
        \includegraphics[width=0.95\linewidth]{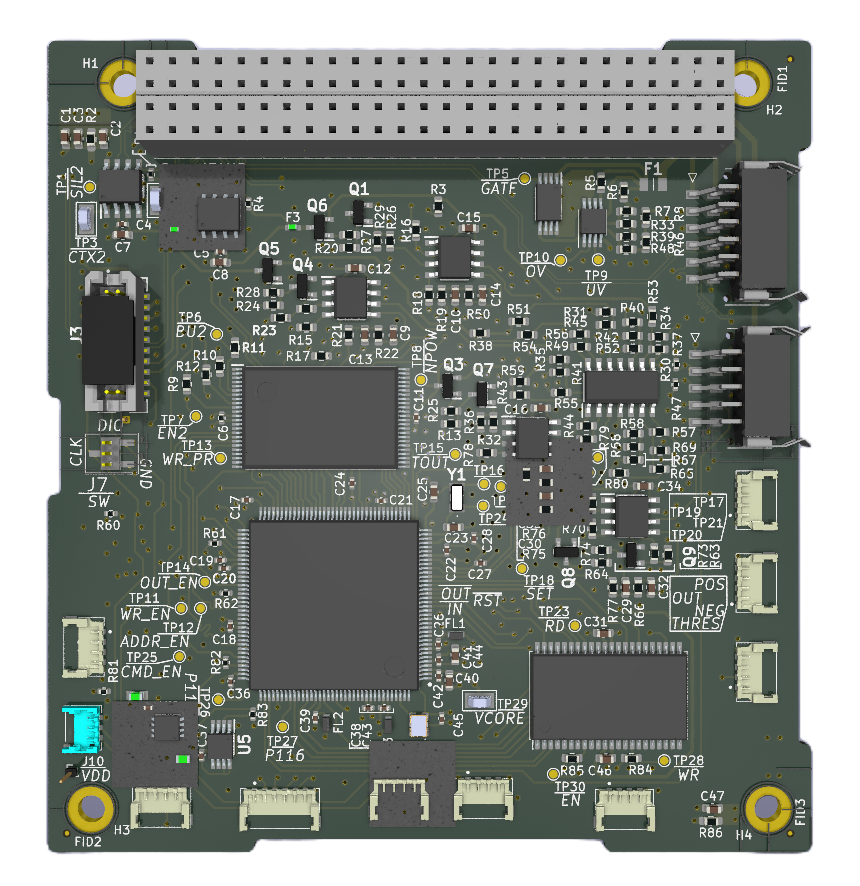}
        \caption{\acs{OBC} side, top layer}
        \label{fig:obc-board-top-layer}
    \end{subfigure}
    \hfill
    \begin{subfigure}{0.495\linewidth}
        \centering
        \includegraphics[width=0.95\linewidth]{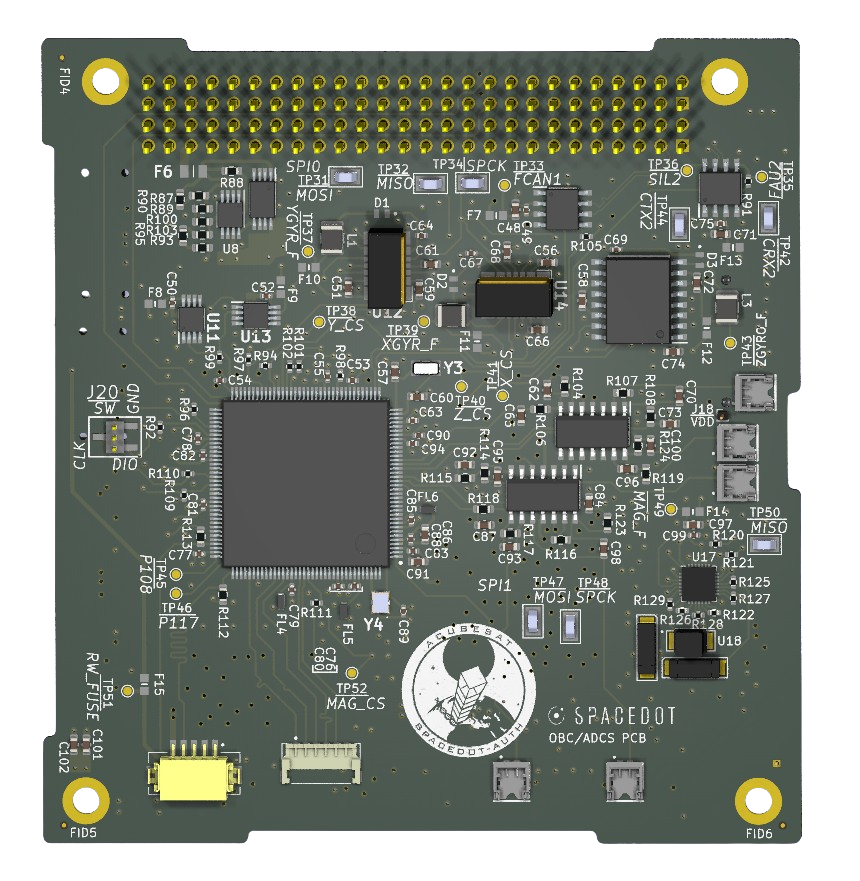}
        \caption{ADCS side, bottom layer}
        \label{fig:adcs-board-bottom-layer}
    \end{subfigure}
    \caption{\acs{OBC}/\acs{ADCS} board render with all components layed out. The top layer hosts the \acs{OBC} subsystem (a) and the bottom layer the \acs{ADCS} (b).}
    \label{fig:board-cad}
\end{figure*}

At the hardware implementation level, a single \acs{PCB}, that follows the LibreCube Board Specification \cite{LibreCube}, seen in \Cref{fig:board-cad}, hosts two distinct subsystems. One resembling a conventional \acs{OBC} responsible
for telecommand execution, telemetry retrieval, on-board time synchronization, in-orbit patching \cite{DDJF_OBDH}, and system-level \acs{FDIR} \cite{FMEA}. The
second subsystem entails the implementation of an \acs{ADCS} for both nominal and downlinking scenarios \cite{DDJF_OBDH,DDJF_AOCS}. The two subsystems have their own dedicated SAMV71Q21RT \acs{MCU}, for a more distributed processing power and reliability. The decision to combine both subsystems onto a single \acs{PCB} was driven by space constraints. From this point forth, the \acs{PCB} hosting the two subsystems will be referred to as the \acs{OBC}/\acs{ADCS} board, with each subsystem referred to as the \acs{OBC} side and the \acs{ADCS} side, respectively.

Software-wise, an early decision was taken to extensively use modern C++17 for the development of the system firmware \cite{DDJF_OBSW}, in order to exploit the extensive compile-time type safety checks \cite{CPP-Core-Guidelines} provided by the language standard. The team deliberately eschewed features such as dynamic memory allocation, multiple inheritance, virtual functions, and the default \acf{STL}, preferring to replace it with an embedded-friendly implementation, namely \acs{ETL} \cite{ETL}, in order to reduce the binary size and make memory allocation behaviors more predictable. As the complex nature of such a control software requires multiple tasks that need to run in parallel, FreeRTOS \cite{FreeRTOS} was selected as the operating system as it natively supports the \acs{ARM} Cortex-M7 architecture which is the core of the SAMV71Q21RT \acs{MCU} \cite{FreeRTOS_supported_devices}. The software source files can be found in the project's open repositories for \acs{OBC} \cite{OBC-EQM-SW-Gitlab} and \acs{ADCS} \cite{ADCS-EQM-SW-Gitlab}.

The system level design philosophy consists of the \acf{IM} (flatsat) and the \acf{PFM} (assembled spacecraft) \cite{MAIVP}. Every procured subsystem component initially integrated into the \acs{IM}. Specifically for the \acs{OBC}/\acs{ADCS} board, the design approach includes the production of a \acf{DM}, an \acf{EQM} and finally a \acf{FM} \cite{MAIVP}.

\subsection{System Architecture}
\label{sec:obc-architecture}

The \acs{OBC} is responsible for the reliable management of information flow throughout the entire spacecraft's system \cite{DDJF_SYS} and it's built around the hardware specification seen in \Cref{tab:hardware-specs-obc}.

\begin{figure}[!ht]
    \centering
    \includegraphics[width=0.95\linewidth]{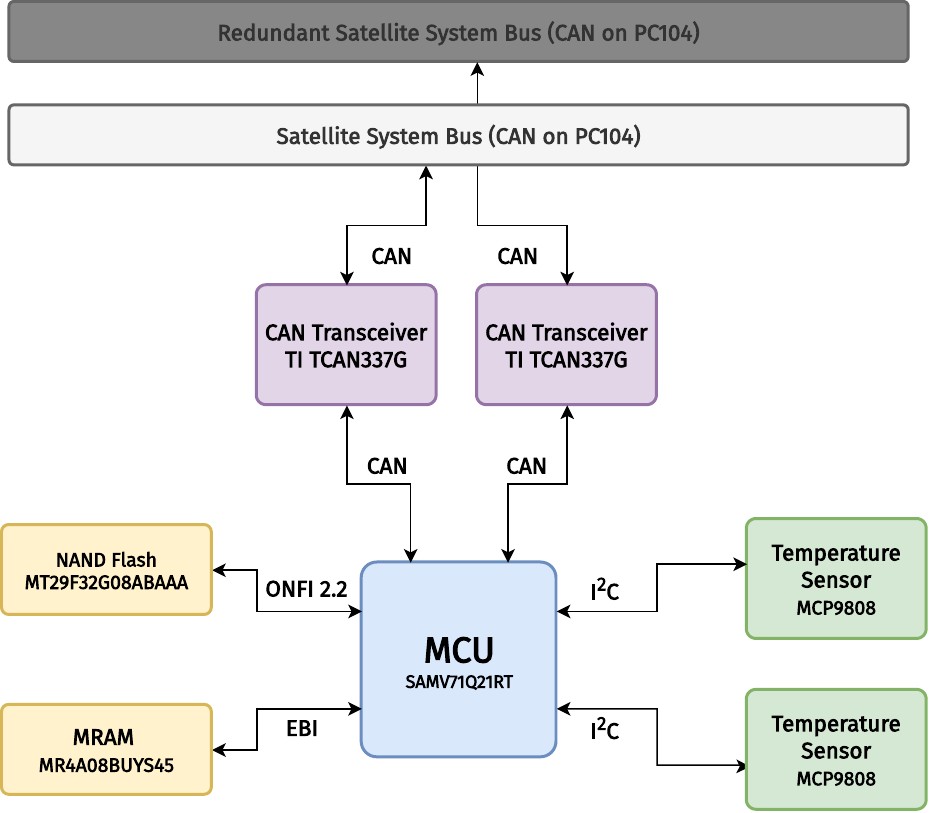}
    \caption{High level diagram of the \acs{OBC} board physical architecture. The letter and number sequence below the names in each block represents the part number of the respective component.}
    \label{fig:obc-physical-arch-high-level}
\end{figure}

The chosen \acs{MCU} has a drop-in pin-compatible replacement non-tolerant version that is used for development purposes, thus reducing production costs for experimentation \cite{ATSAMV71Q21B-AAB}. Data storage for mission-critical parameters is supported by \SI{2}{\mega\byte} of \acf{MRAM} \cite{MR4A08BUYS45}, selected for its inherent immunity to \acf{SEU}s \cite{MRAM_radiation_tolerance}, while non-critical data (e.g. telemetry) are stored in a \SI{4}{\giga\byte} automotive-grade \acf{NAND} Flash \cite{NAND_Flash}. A brief high-level diagram of the \acs{OBC} subsystem architecture is depicted in \Cref{fig:obc-physical-arch-high-level}.

Given the satellite’s prolonged exposure to ionizing radiation, safeguarding critical electronics is essential. To enhance their resilience in these harsh conditions, the team developed custom in-house \acs{LCL} circuits. These circuits were specifically designed to protect the most vital components of the \acs{OBC}, the two memory modules and the \acf{CAN} transceivers, from \acf{SEL}.\label{sec:radiation-first-mention}

 The functions of the \acs{ADCS} subsystem serve to control the spacecraft's angular velocity within safe limits, pointing during payload data transmission, in addition to sun pointing for maximal power input. To achieve that, the \acs{MCU} is connected to a series of sensors that aid in the estimation of the spacecraft's orientation and angular velocity, driving the required actuators for control \cite{DDJF_AOCS}. The main components that constitute the \acs{ADCS} can be found in \Cref{tab:hardware-specs-adcs}.

All satellite \acs{PCB}s, excluding the Payload Subsystem, are connected through a stack configuration using LibreCube-compatible PC104 connectors \cite{LibreCube}. The 104-pin connectors provide flexibility, ensuring compatibility with various Open-Source and \acf{COTS} hardware modules. Communication between the \acs{OBC} and \acs{ADCS} subsystems, as well as with other spacecraft systems, is facilitated through the \acf{CAN-FD} bus, which operates with a custom signaling protocol \cite{DDJF_OBDH} that transports messages according to the \acs{ECSS} \acf{PUS} standard \cite{PUS}. The bus is cold-redundant, ensuring inter-subsystem communication in case of failure. The SatNOGS Communications subsystem \cite{SatNOGS-COMMS}, which is part of the satellite, is fully compatible with this design. \acs{CAN-FD}, along with the transceiver specified in \Cref{tab:hardware-specs-obc}, was selected for compatibility with the SatNOGS but also for its built-in protection features, including overvoltage and \acf{ESD} safeguards \cite{TCAN33x}.

\begin{table*}[!hp]
    \centering
    \caption{Subsystem hardware specification of the \acs{OBC} and \acs{ADCS}. The first column in each table shows the component's function and the second column shows the details, like the part number.}
    \label{tab:hardware-specs}
    \begin{subtable}[t]{0.495\linewidth}
        \centering
        \caption{\acs{OBC} hardware specification}
        \begin{tabularx}{\linewidth}{lX}
            \hline
            \textbf{\acs{MCU}} & SAMV71Q21RT, (ARM Cortex-M7 @ \SI{300}{\mega\hertz}) \cite{SAMV71Q21RT} \\ \hline
            \textbf{Spacecraft bus} & 2x Texas Instruments TCAN337G CAN Transceiver \cite{TCAN33x}\\ \hline
            \textbf{Critical data storage} & Everspin MR4A08BUYS45 \acs{MRAM}, \SI{2}{\mega\byte} \cite{MR4A08BUYS45}\\ \hline
            \textbf{Ambient sensing} & 2x Microchip MCP9808 Temperature Sensors \cite{mcp9808}\\ \hline
            \textbf{Main storage} & Micron MT29F32G08ABAA NAND Flash, \SI{4}{\giga\byte}, \acs{SEE} data available \cite{NAND_Flash} \\ \hline
            \textbf{Active \acs{SEL} countermeasures} & In-house-designed \acsp{LCL} \\ \hline
            \textbf{Protection Circuit} & Overvoltage, Undervoltage and Reverse Surge Protector LTC4367 \cite{OV-UV_Protection}\\ \hline
        \end{tabularx}
        \label{tab:hardware-specs-obc}
    \end{subtable}
    \hfill
    \begin{subtable}[t]{0.495\linewidth}
        \centering
        \caption{\acs{ADCS} hardware specification}
        \begin{tabularx}{\linewidth}{lX}
            \hline
            \textbf{\acs{MCU}} & SAMV71Q21RT, (ARM Cortex-M7 @ \SI{300}{\mega\hertz}) \cite{SAMV71Q21RT} \\ \hline
            \textbf{Spacecraft bus} & 2x Texas Instruments TCAN337G CAN Transceiver \cite{TCAN33x}\\ \hline
            \textbf{Sun-Spacecraft Orientation} & Coarse Sun Sensors \\ \hline
            \textbf{Angular Acceleration} & Magnetorquers ISIS iMTQ \cite{MTQ}\\ \hline
            \textbf{Ambient sensing} & 2x Microchip MCP9808 Temperature Sensors \cite{mcp9808} \\ \hline
            \textbf{Angular Acceleration} & Reaction Wheel RW210 \cite{RW}\\ \hline
            \textbf{Angular sensing} & 3x Gyroscopes ADXRS453BEYZ (X,Y,Z Axes) \cite{Gyroscope}\\ \hline
            \textbf{Earth magnetic field sensing} & 3-axes Magnetometer RM3100 \cite{Magnetometer}\\ \hline 
            \textbf{Protection Circuit} & Overvoltage, Undervoltage and Reverse Surge Protector LTC4367 \cite{OV-UV_Protection}\\ 
            \hline
            \end{tabularx}
        \label{tab:hardware-specs-adcs}
    \end{subtable}
\end{table*}

\subsection{Manufacturing}
\label{sec:design-and-manufacturing}

The \acs{OBC}/\acs{ADCS} \acs{PCB} was entirely designed using the KiCad open-source EDA tool \cite{KiCad}, with the design files available in the project's GitLab repositories \cite{OBC-ADCS-Board-Gitlab}. The PCB features an 8-layer design, including two layers dedicated to components and signal routing, three for ground planes, and one for power distribution. The components, corresponding to the \acs{OBC} and \acs{ADCS} subsystems, are placed on opposite sides of the \acs{PCB} and are electrically isolated through the ground layers to prevent interference. The multi-layer structure also provides additional resilience against extreme vibrations.

The \acs{PCB} was manufactured by Eurocircuits and assembled by \textbf{Prisma Electronics S.A}\footnote{\href{https://www.prismaelectronics.eu/index.php/en/}{https://www.prismaelectronics.eu}}. The solder and flux types used are Sn63Pb37, Type K/3.5\% ROL0, specifically SAC305 and PF7723, according to the \acs{ECSS} soldering standards \cite{ECSS-FLUX}, optimized for Space applications to ensure robustness under extreme temperature conditions.

\subsection{Subsystem Space Validation}
\label{sec:space-validation}

Given the hostile conditions of the Space environment, it is essential to expose satellite modules to similar conditions early in the development process to validate their performance. As mentioned earlier in \Cref{sec:radiation-first-mention}, the radiation levels in \acs{LEO} necessitate testing the radiation tolerance of key components. To address this, a dedicated radiation testing campaign was conducted at the Heavy Ion Testing facility at UCLouvain, as detailed in \Cref{sec:radiation}.

In addition to radiation tolerance, the \acs{PCB} must endure the mechanical stresses of the launch and extreme temperature effects. \acf{TVAC} testing, along with vibration tests, play a critical role in ensuring the board’s durability and overall mission success. The team performed two environmental test campaigns at \acs{ESA}’s CubeSat Support Facility at \acf{ESEC}-Galaxia in Belgium. These tests provided valuable insight in the functionality of the board, which is discussed in detail in \Cref{campaigns}.

Currently, the team is actively refining both the hardware and software of the \acs{OBC}/\acs{ADCS} board, drawing on insights from these environmental tests and ongoing assessments under ambient conditions. Debugging and iterative development are helping address the challenges identified during testing.

The following sections present the key results of these testing campaigns, along with the lessons learned and the troubleshooting efforts undertaken to enhance the system’s robustness.

\section{Radiation Effect Mitigation}
\label{sec:radiation}

To enhance the reliability of the \acs{OBC}/\acs{ADCS} board, the team adopted a specific design approach focused on mitigating radiation effects. Protective measures and redundant circuitry following the identification of the critical systems have been implemented (as mentioned in \Cref{sec:obc-architecture}). The team opted to use \acs{COTS} components, aligning with the New Space industry trend, a decision that also significantly reduced the costs of the in-house \acs{OBC}/\acs{ADCS} board models.

The component selection followed a common pipeline of creating a database and gathering existing information online, with respect to \acf{TID} and \acf{SEE} radiation immunity levels. The approach taken was the addition of a radiation design margin of at least 1.2 (as mentioned in the \acs{ECSS} policy for design margins \cite{ECSS-RADIATION}) to account for any uncertainties concerning the simulated values, as well as, any extra degradation, rendering components more prone to radiation effects. It should be noted that the \acf{TID} levels were simulated with the TRAD's FASTRAD \cite{FASTRAD} simulation tool by using the full \acf{CAD} of the satellite and the orbital data as input \cite{MDO}. The simulated levels for the mission duration were found to be well within the \acf{RDM} limits. To increase the reliability and tackle possible issues connected to the STM micro-controllers (SEE sensitivity), the team decided to use a radiation-tolerant \acs{MCU} for the \acs{FM} and the equivalent model non-radiation hardened version for the \acs{EQM}.


Certain components identified as critical, with missing radiation-related data, were screened for \acs{SEE} effects during the Heavy Ion Testing at UCLouvain. The main focus of the testing was on \acs{SEL}s, given their potentially destructive nature \cite{ECSS-RADIATION}. Before testing the components, the protective cover was removed through a process called delidding. With the help of the EEE (Electrical, Electronic and Electro-mechanical) components lab of ESTEC \cite{MECL}, part of the casing was eliminated so that the die could be directly exposed to the Heavy-Ion beam. After the delidding process, the components were soldered on a dedicated testing board.

\begin{figure}[h!]
    \centering
    \includegraphics[width=0.95\linewidth]{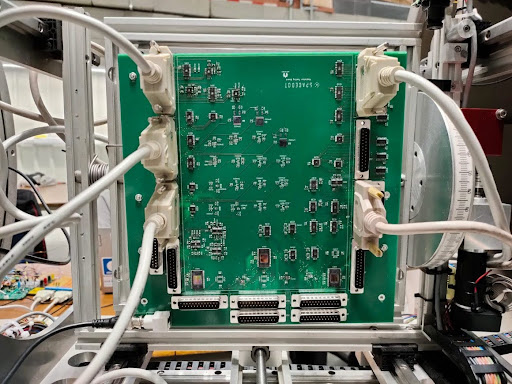}
    \caption{Heavy ion testing board with all the components soldered}
    \label{fig:radtest}
\end{figure}

The initial list of components identified included the \acs{MRAM} \cite{MR4A08BUYS45}, the CAN Transceiver \cite{TCAN33x}, the 555 timer (TLC555QDRQ1) \cite{TLC555}, the shift register (SN74HC595DR) \cite{SNx4HC595}, the Operational Amplifier (TLV2254AIDR) \cite{TLV225}, the NOR Gate (CD74HC02ME4) \cite{CDx4HC02} and the S-R Latch (CD4044BD) \cite{CD4043}.


The final version of the in-house \acs{LCL}-hosted parts that were tested in the \acf{HI} campaign. The \acs{LCL} development board was also used during the campaign as part of the \acf{EGSE}, resulting in both the validation of the functionality of the board and the qualification of the individual components.
The key takeaway from this testing campaign was the absence of SELs for every tested component. Due to time constraints and hardware issues, the memories were not tested. \acs{CAN} transceivers were also tested for a short period of time, during which, Single Event Transients (\acs{SET}s) and Single Event Upsets (\acs{SEU}s) were observed.




\section{Environmental Testing Campaigns}
\label{campaigns}
The two Environmental Testing Campaigns with a single \acs{EQM} model of the AcubeSAT \acs{OBC}/\acs{ADCS} board design serving as the \acf{DUT} were conducted at the ESA \acs{CSF}, ESEC-Galaxia, Transinne, Belgium \cite{CSFWebsite}. The campaign consisted of Vibration and \acs{TVAC} testing. The detailed Test Specifications and Test Reports can be found in the relevant \acf{TSTP} documents \cite{OBCTVAC1,OBCVIBE1,OBCTVAC2,OBCVIBE2}.

\subsection{Tests Configuration}
\label{set-up}

\subsubsection{Vibration Testing Description}
In accordance with the "Test as you Fly, Fly as you Test" testing principle, vibration testing took place before \acs{TVAC} tests. During the vibration testing the \acs{DUT} underwent on each axis a Resonance Search to identify the natural frequencies of the \acs{DUT}, followed by a Random Vibration test and a post Resonance Search to verify structural integrity by observing minimal frequency and amplitude shifts in the frequency response.
Three main measurement accelerometers were used. Two of them were placed on the \acs{ADCS} side, one on top of the \acs{MCU} where the maximum effects of the loading are expected, and one near the corner of the \acs{PCB}. The third accelerometer was placed on the metal adaptor used for the mounting on the shaker, along with two control accelerometers.

\begin{figure*}[!ht]
    \centering
    \begin{subfigure}{0.495\linewidth}
        \centering
        \includegraphics[width=0.75\linewidth]{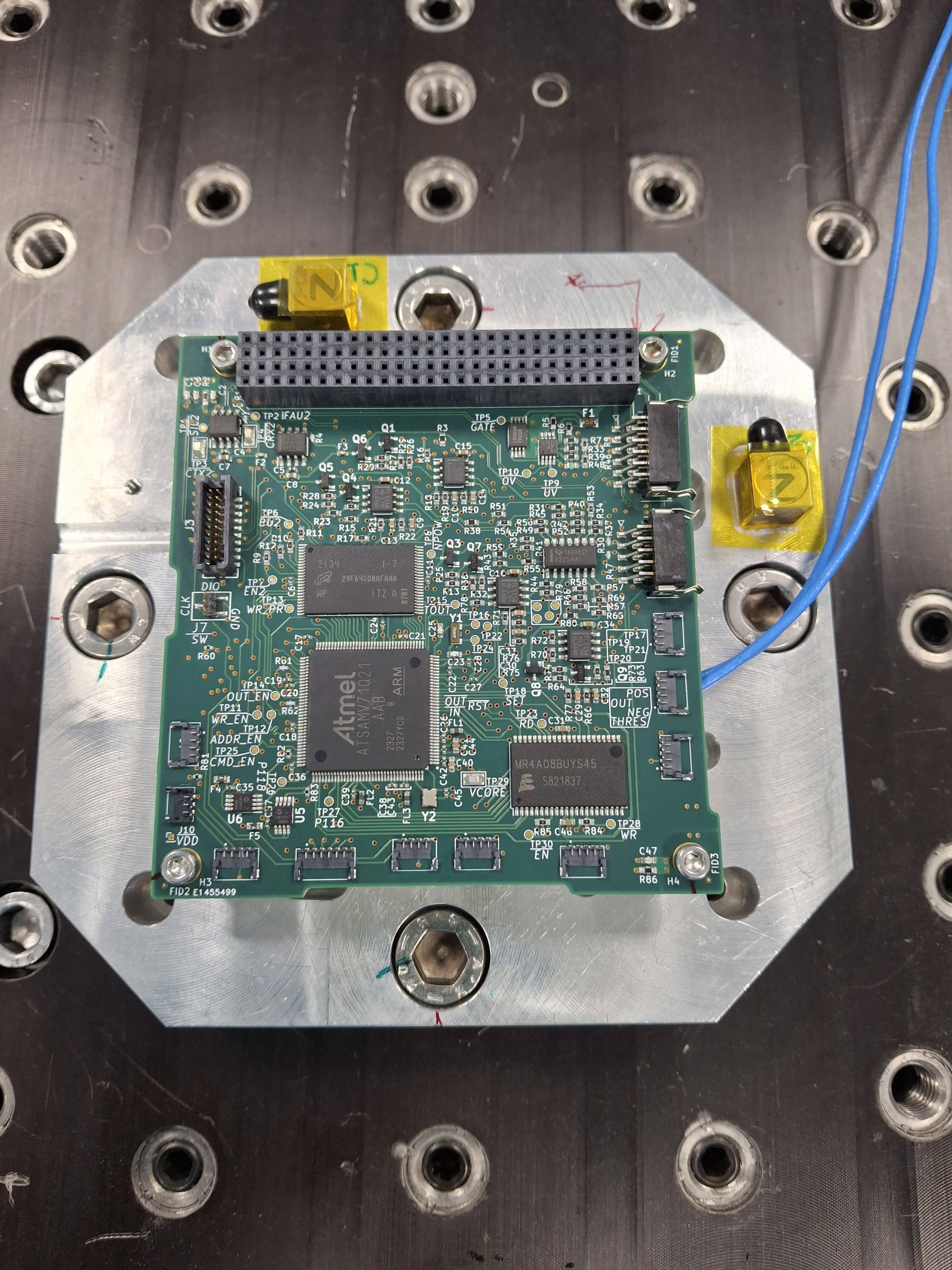}
        \caption{\acs{VIBE} Armature, \acs{OBC} side up.}
        \label{fig:board-on-armature}
    \end{subfigure}
    \hfill
    \begin{subfigure}{0.495\linewidth}
        \centering
        \includegraphics[width=\linewidth,  angle = 270]{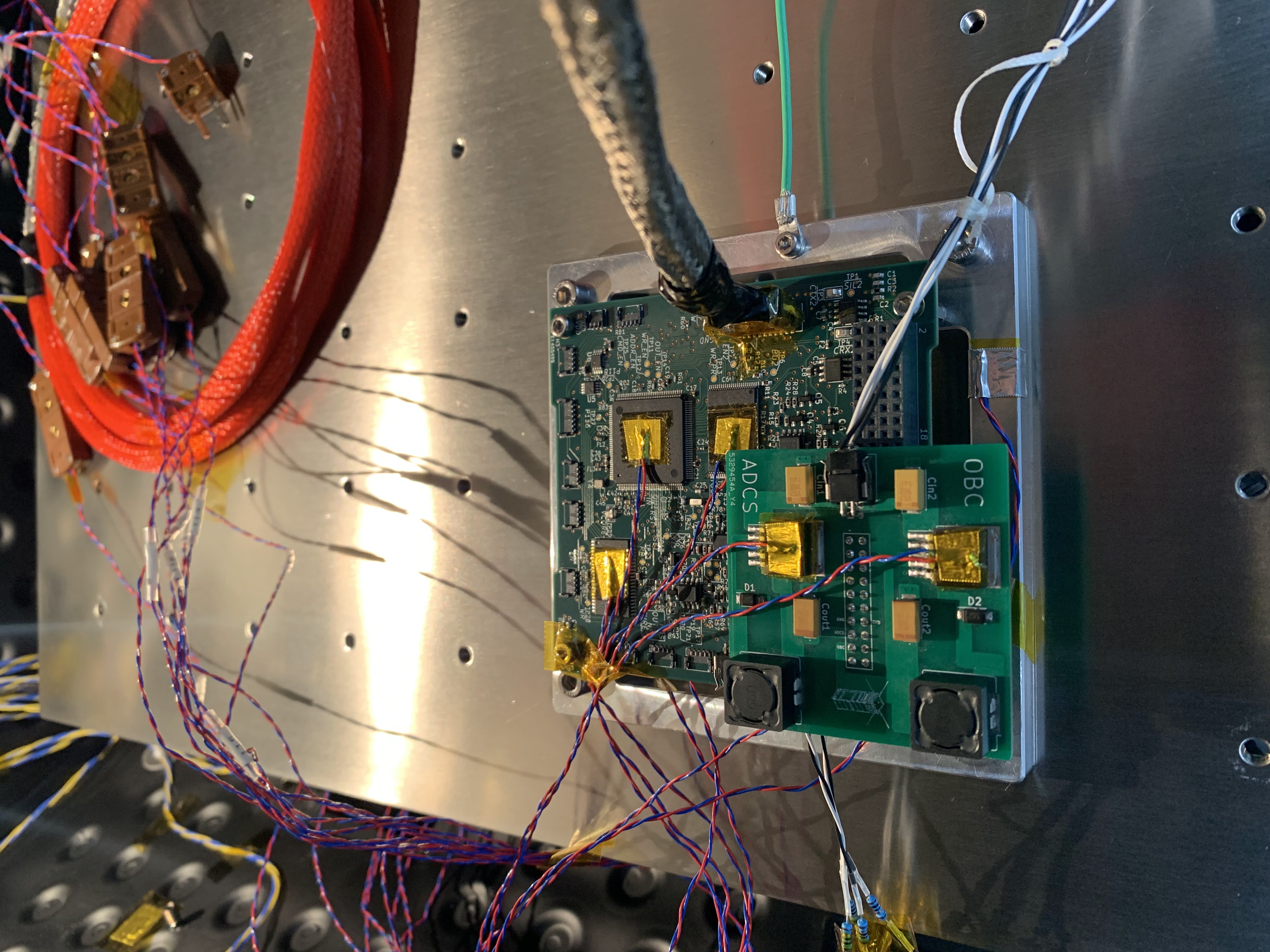}
        \caption{Board in Thermal Vacuum Chamber}
        \label{fig:tvac-placement}
    \end{subfigure}
    \caption{\acs{OBC}/\acs{ADCS} board with all sensors attached for the \acs{VIBE} and \acs{TVAC} tests.}
    \label{fig:tvac-vibe}
\end{figure*}

\subsubsection{TVAC Testing Description}
The \acs{TVAC} testing includes four thermal cycles due to time constraints. One is non-operational, where the \acs{DUT} is turned off, and three are operational where the \acs{DUT} must withstand temperature fluctuations while being operational. Its functionality is confirmed through functional tests, either at the maximum temperature point or correspondingly at the minimum.
The placement of the thermocouples is designed to capture the thermal response of essential components on both the \acs{OBC} and \acs{ADCS} sides of the board and also takes into account the need for adequate surface area to securely tape down both the thermocouples and the associated cables. 

\subsection{Environmental Testing Requirements}
\label{campaigns-requirements}
The preparation and execution of an environmental testing campaign for the \acs{OBC}/\acs{ADCS} board must adhere to the system and subsystem requirements outlined in the \acf{VCD}s \cite{TSVCD,FDSVCD}. The test campaign ensures the board’s capability to function reliably in space conditions by verifying compliance with performance, equipment, and environmental standards.
Additionally, the testing facilities must meet CubeSat team standards, such as using appropriate thermal and vibration testing equipment under controlled conditions. Moreover, the unique demands of each environmental test—such as ensuring the board's functionality is not impacted are verified through functional tests.
The campaign requirements are specified in the relevant \acs{TSTP} documents \cite{OBCTVAC1,OBCVIBE1,OBCTVAC2,OBCVIBE2}.

\subsubsection{Vibration Testing Requirements}
\label{campaigns-vibe-requirements}

The vibration testing requirements aim to ensure that the \acs{PCB} can endure mechanical stresses without functional degradation, with a particular focus on assessing its structural integrity and preventing damage. The philosophy also prioritizes careful monitoring through resonant searches and visual inspections to detect any potential weaknesses. Additionally, precision in the placement and mounting of measurement tools, such as accelerometers, is critical to maintaining the reliability and accuracy of the test results.

\subsubsection{Thermal Vacuum Testing Requirements}
\label{campaigns-tvac-requirements}

The \acs{TVAC} requirements are driven by the philosophy of replicating the harsh conditions of space to ensure that the board can perform reliably in extreme thermal and vacuum environments. There is an emphasis on maintaining functional integrity across a wide temperature range and ensuring that the equipment can operate under stressful thermal conditions.

\subsection{First Campaign}
\label{first-campaign}

\subsubsection{Vibration Testing Results}
\label{first-vibe}
During all sequences, there were amplitude shifts in the pre and post-random Resonance searches. The amplitude shift on the X axis was attributed to the natural settling of the \acs{DUT}, the other shifts were negligible as elaborated in \cite{OBC_EQM_TRPT_1}. After the tests, the \acs{DUT} functionality remained unimpaired. As a result, the testing process was deemed successful. The pass-fail criteria specified were all met for the tests conducted and the team considers the \acs{OBC}/\acs{ADCS} board \acf{VIBE} qualified.


\subsubsection{Thermal Vacuum Testing Results}
\label{first-tvac}
During \acs{TVAC} testing, after pumping down and before performing any thermal cycling, it was observed that the \acs{OBC} \acs{MCU} kept resetting at a random period and the \acs{MCU} was also informed about \acf{CAN} bus errors. This anomaly occurred at approximately 40 $^\circ$C and functionality recovered after the temperature dropped.

After software debugging at the \acs{CSF} \cite{OBC_EQM_TRPT_1} during a high-temperature stimulation with a heat lamp, the cause of the crash was determined to be the communication between the \acs{MCU} and the \acs{MRAM} along with the overclocking of the \acs{MCU} due to the external 16 MHz \acf{TCXO} that was outputting a frequency of 304 MHz for the CPU clock instead of the nominal 300 MHz. This was also causing the CAN errors.  

When the board was subjected to sub-zero temperatures, none of the measurements were negative and, instead, had an offset of around 30 degrees over the chamber's temperature, indicating an issue in the logic concerning negative values included in the implementation of the driver. Furthermore, the \acs{ADCS} \acs{MCU} experienced a circumstantial crash, occurring at -20 $^\circ$C but was reset by the internal watchdog and continued to function normally. During the second cold cycle, the team attempted to underclock the \acs{MCU} due to the use of a \acs{TCXO} crystal that differed from the one intended for the \acs{FM}. Although the issue reoccurred, the \acs{MCU} failed to recover and remained in an idle state. It is important to note that while the board was in the \acs{TVAC}, the team was unable to debug or upload code due to \acf{GSE} issues, which limited the ability to troubleshoot in real-time. 

Following the TVAC test, all electronics and circuitry remained operational as verified through the \acs{FT}s. The Product Assurance aspect of \acs{TVAC} testing is considered successful and the manufacturing and assembly processes are considered to be verified. Nevertheless, functionality in extreme temperatures and vacuum is ultimately not reliable and the  \acs{OBC}/\acs{ADCS} board is not \acs{TVAC} qualified.

\subsection{Design Improvements Based on First Campaign}
After the first campaign, four \acf{NCR}s were filled, indicating problems with the temperature sensors software \cite{NCR-TEMP}, \acs{MCU} crashes at high temperatures \cite{NCR-HOT}, communication failures via the \acs{CAN} Bus at elevated temperatures \cite{NCR-CAN}, and the unresponsiveness of the \acs{ADCS} \acs{MCU} at low temperatures \cite{NCR-COLD}. Each of these issues was systematically addressed through targeted design modifications, which were subsequently validated through rigorous testing.

\subsubsection{Temperature Sensor Software Malfunctions}
The driver caused the external temperature sensors to show unreliable temperature values. More specifically, when the temperature went below 0 $^\circ$C, none of the measurements were negative and, instead, had an offset of around 30 $^\circ$C over the chamber's temperature. The error was induced by a fault in the logic concerning negative values included in the implementation of the driver. Modifications in the driver's software successfully resolved the issue as verified by subsequent tests using an MCP9808 \cite{mcp9808} evaluation board.

\subsubsection{\acs{MCU} Crashes at High Temperatures}
A significant issue identified during testing was the recurrent crashing of the \acs{OBC} \acs{MCU} under elevated thermal conditions. The analysis in the \acs{CSF} revealed that this instability derived from the use of an \acf{RC} circuit-based clock, which lacked sufficient thermal stability, the overclocking of the \acs{MCU} due to the higher frequency \acs{TCXO} and the \acs{MRAM}'s operations. To mitigate this, the \acs{RC} circuit-based clock was replaced with a 12 MHz \acs{TCXO} which offers greater reliability in extreme temperatures and produces an expected output frequency to the \acs{MCU} clock. In addition, the \acs{MRAM} read and write timing cycles were optimized via \acf{SMC} configuration.\\
The performance of the system following these changes improved markedly. The \acs{MCU} remained operational up to 49 $^\circ$C, significantly extending its tolerance to high temperatures. Furthermore, isolation heat tests confirmed that the failure was component-related and originated from the \acs{MCU}, as heating the \acs{MCU} alone caused it to fail at an internal temperature of 67°C and an ambient of 46 $^\circ$C, while the \acs{MRAM} and other components remained stable for temperatures above 50 $^\circ$C. Subsequent heat tests on another breakout module and an \acs{OBC} \acs{EM} board, both having similar functionality and architecture, showed stable performance. Upon replacing the defective \acs{OBC} \acs{MCU} on the \acs{OBC}/\acs{ADCS} Board, the issue was fully resolved, with no further crashes observed during the ensuing thermal testing.

\subsubsection{CAN Bus Failures}

Errors were occurring on both the \acs{OBC} and \acs{ADCS} during communication over the \acs{CAN} Bus when the temperature was over 52 $^\circ$C. In either case, as the temperature further deviated from ambient, communication was automatically halted as neither side could successfully transmit or receive messages. The team was forced to use the internal \acs{RC} clocks on both \acs{MCU}s since the \acs{TCXO} crystal differed from the one that was intended to be used, due to supply chain issues at the time of the \acs{EQM} manufacturing. After replacing the \acs{TCXO} crystal to match the \acs{FM} configuration, the clock output was within the datasheet's \cite{TCXO_Datasheet} specification of a 2.5 ppm tolerance anywhere in the operating range from -40 $^\circ$C to +85 $^\circ$C. the communication integrity was restored, with the \acs{CAN} Bus functioning reliably during high-temperature tests. 

\subsubsection{\acs{ADCS} \acs{MCU} Unresponsiveness at Low Temperatures}

At low temperatures, the \acs{ADCS} \acs{MCU} became unresponsive, which compromised system functionality during cold environment tests. While a certain cause has not been identified yet, the team suspected that the RC circuit-based clock led to the initial crash during the first cold cycle and the \acs{MCU} complete unresponsiveness emerged due to
the underclocking of the microcontroller while using the wrong model of the temperature-compensated
crystal, during the second cold cycle. However, a failure of the Central Processing Unit could also occur from the manufacturing process resulting
in a faulty component. Due to the high risk of damaging the board with cold tests outside the \acs{TVAC} environment, caused by potential condensation from humidity, no further troubleshooting was conducted.



Given these changes and after thorough testing, having verified the board's stable behavior at hot temperatures, the team decided to repeat the environmental testing to verify that the solutions were effective.

\subsection{Second Campaign}
\label{second-campaign}

\subsubsection{Vibration Testing Results}
\label{second-vibe}
The \acs{OBC}/\acs{ADCS} board encountered significant issues during the \acs{VIBE} testing leading to a failure of the qualification process. The testing followed a standard sequence across all three axes (X, Y, and Z), and while initial results were in line with expectations, a major anomaly occurred during the random sequence. A large amplitude shift of 20\% was noted, which was attributed to the natural settling of the assembly. However, subsequent tests showed amplitude shifts reaching 481\%, indicating a serious issue with the integrity of the \acs{DUT}.

Upon thorough inspection \cite{OBC_EQM_TRPT_2} after the Z-axis testing, it was discovered that two critical inductors, responsible for magnetometer measurements on the X and Z axes, had detached from the \acs{PCB} \cite{NCR-VIBE}. This detachment compromised the data acquisition from the magnetometer, rendering the results invalid \cite{NCR-Magnetometers}. As a result, the functional tests post-\acs{VIBE} were conducted with limited functionality.

\subsubsection{Thermal Vacuum Testing Results}
\label{second-tvac}
The \acs{TVAC} testing of the \acs{DUT} encountered a few critical issues, though some components, like the \acs{NAND} Flash, the \acs{MRAM}, and the \acs{CAN} Bus of the system performed successfully. During the test, random crashes of both the \acs{OBC} and the \acs{ADCS} \acs{MCU}s occurred intermittently \cite{NCR-Crash-TVAC}, with the crashes becoming more frequent during transitions from hot to cold cycles. After re-flashing the \acs{OBC}'s \acs{MCU}, during the cold temperature dwell, the board stabilized, but the \acs{ADCS}'s \acs{MCU} consistently crashed under low temperatures and during transitions. Through software debugging \cite{OBC_EQM_TRPT_2} potential scheduling conflicts and hardware malfunctions were identified as causes. These issues, including potential \acs{GSE} reliability and temperature-induced malfunctions, led to the conclusion the \acs{TVAC} testing of the \acs{OBC}/\acs{ADCS} board was unsuccessful.

A post-\acs{TVAC} inspection revealed no visible damage or deformations to the \acs{OBC}, \acs{ADCS}, or the voltage regulator \acs{PCB} supplying the \acs{EQM}. No permanent damage was sustained, so additional functional tests were deemed unnecessary.

\section{Current Status}
\label{current-status}
At present, the status of the board indicates that all systems are operational in ambient conditions and the board exhibits no critical faults. Despite that, the aforementioned issues related to the \acs{MCU}s' unexpected behavior under certain temperature ranges remain under investigation.   
Following the second environmental testing campaign, more extensive testing of the board was required to identify the issues. The reproduction of the unstable behavior was not accomplished during a following thermal cycle that was conducted on the same board, using the exact same ground support equipment that was utilized during the official environmental qualification campaigns. Regarding the specifications of this test, the temperature range of 50°C to -30°C, which was considered adequate for the temperature range that the issues were noticed initially, was achieved for a shorter duration than the thermal cycles during the two campaigns, with a lower temperature gradient and using a thermal chamber, without the effect of the vacuum conditions.

As of the latest testing phase, after a thorough inspection of the \acs{TVAC} Umbilical cable, it was observed that when bending the cable in different ways the functionality of the board was affected, resulting in the unstable behavior of the board both in ambient and in testing conditions. Another significant issue that resulted in the crashes of the \acs{OBC} \acs{MCU}, was identified to be the driver of the \acs{MRAM}, which needs to be further optimised to meet critical timing constraints and edge cases during temperature transitions.

Following the replacement of the umbilical cable to eliminate any signal degradation, along with necessary software adjustments, the \acs{OBC} \acs{MCU} demonstrates full operational capacity under all conditions tested within the thermal chamber. 

Attempts to reproduce the \acs{ADCS} \acs{MCU} crash under similar conditions, outside a \acs{TVAC}, have so far been unsuccessful, leaving it as the sole unresolved issue.

\section{Future Work}
\label{future-work}

Further investigation is required to reproduce the issues and fully understand the root cause of the \acs{MCU}s' unstable performance, under specific environmental conditions. After the identification of the causes, a 3rd Environmental Testing Campaign will be pursued with a new manufactured model that will comply with any possible required changes, induced by the successful resolution of the problem.

Upon completing the prerequisite Space qualification, conducted and radiated emission \acs{EMC} testing will be carried out, avoiding any destructive tests. With the characterization of the components of the \acs{PCB} already completed, the next step will be the implementation of a Test Plan.

On the firmware side, optimization and finalization of the on-board software, including memory management, \acs{ECSS} services implementation, and subsystem-level \acs{FDIR} are included in the required work for the \acs{OBC}. The implementation of these is considered necessary to ensuring the data handling and timing requirements related to the simultaneous control of the various subsystems of the satellite, thus minimizing the risk of system conflicts or communication breakdowns. On the \acs{ADCS} side, the pending work comprises the finalization of the actuators' drivers and the integration of the determination and control algorithms in the onboard software.

Following the subsystem-level testing and verification, the integration process will be initiated. At an early phase, using a flatsat configuration as described in AcubeSAT's \acf{MAIVP} \cite{MAIVP}, the communication of the \acs{OBC} with the rest of the subsystems of the satellite will be tested and established. More specifically, the \acs{OBC} will interface with the \acs{COMMS} board, the \acs{EPS}, the \acs{ADCS} (including the \acs{ADCS} \acs{MCU}, the Reaction Wheel, the Magnetorquers and the photodiodes), the Antenna Deployment Mechanism, the Payload and the Solar Panels. Once the integration is complete hardware-wise, the remaining system-level software is to be finalized and the implementation of the switching of the operational modes of the satellite, \acs{FDIR}, in orbit patching and packet storage is going to be tested.

\section{Conclusion}

The journey of developing AcubeSAT’s \acs{OBC}/\acs{ADCS} board has been one of continuous iterations and learning. Through extensive testing, the team has refined the design, solved critical issues, and improved the board’s overall reliability. Each step has revealed important lessons that guided design improvements and provided the foundation for future testing. Looking forward, the team aims to finalize the software and hardware integration, while addressing remaining thermal challenges through additional testing campaigns, thus developing a subsystem qualified to support the AcubeSAT space mission.


\section{Acknowledgements}

We would like to express our gratitude all the members of the SpaceDot Team for their invaluable help and contribution during the development and testing of the On Board Computer of the AcubeSAT nanosatellite. Specifically, we wish to thank Konstantinos Kanavouras, Athanasios Theocharis for their preliminary work on the \acs{OBC} subsystem, as well as Antonios Keremidis, Christina Koutsou, Xhulio Luli, Dimitrios Koukourikos for their work on the design and testing of the \acs{OBC}/\acs{ADCS} board and everyone that contributed to the subsystem throughout the years. We thank the ESA Education Office and "Fly Your Satellite! 3" Program for the expertise and feedback during the development process and for giving us access to the CubeSat Support Facility at the \acs{ESEC}-Galaxia, Belgium, and sponsoring our travel and accommodation in order to perform both environmental test campaigns. Our warm gratitude extends to the Aristotle University of Thessaloniki for providing us with the proper infrastructure (lab \& cleanroom) and especially the Department of Electrical and Computer Engineering. We also thank Prisma Electronics SA for catering to all our assembly and harness needs. We wish to thank Athanasios Theocharis, Georgios Kikas, Aikaterini Papadopoulou, Stavros Spyridopoulos for providing technical and editorial review. Lastly, we thank our families and friends that supported us all the way by showing patience and understanding.

\AtNextBibliography{\footnotesize}
\printbibliography

\end{document}